\documentclass[12pt]{article}
\usepackage{epsfig}

\begin{document}

\title{Optical Characteristics \\ of the Astrometric Radio Sources}
\author{Zinovy Malkin$^1$, Oleg Titov$^2$}
\date{$^1$ Pulkovo Observatory, St. Petersburg, Russia \\
         $^2$ Geoscience Australia}
\maketitle

\begin{abstract}
A new list of physical characteristics of 3914 astrometric
radio sources, including all 717 ICRF-Ext.2 sources,
observed during IVS and NRAO VCS sessions have been compiled.
The list includes source type, redshift and visual magnitude (if
available). In case of doubt detailed comment is provided. The list
of sources with their positions was taken from the Goddard VLBI
astrometric catalog with addition of two ICRF-Ext.2 sources.  At this
stage the source characteristics were mainly taken from the NASA/IPAC
Extragalactic Database (NED). 667 sources from our list are included
into the IERS list. Comparison has shown a significant difference
in characteristics for about half of these 667 common sources.
We compiled a list of frequently observed sources without known physical
characteristics for urgent optical identification and
spectrophotometric observations with large
optical telescopes. This presented list of physical characteristics
can be used as a supplement material for the ICRF-2, as well as a
database for kinematic studies of the Universe and other related
works, including scheduling of dedicated IVS programs.
\end{abstract}

\vfill
\noindent \hrule width 0.4\textwidth
~\vskip 0.2ex
\noindent {\small 5th IVS General Meeting, St.~Petersburg, Russia, 3--6 March 2007}
\eject

\section{Introduction}

Information on physical characteristics of the geodetic radio sources
is important for planning of VLBI experiments and analysis
of VLBI data to do a research in cosmology, kinematics of the
Universe, etc.  In particular, the primary mainspring to this work
was a support of the investigation of the systematic effects in
apparent proper motion of geodetic radio sources.
\cite{MacMillan05,Titov2007}.

The official list of the physical characteristics of the ICRF radio
sources is supported by the IERS ICRS Product Center (\cite{Archival97}).
The latest version of the IERS list is available in the
Internet\footnote{http://hpiers.obspm.fr/icrs-pc/info/car\_physique\_ext1}.
However this list has some deficiencies:
\begin{itemize}
\itemsep=-0.3ex
\item Not all the sources observed in the framework of geodetic and
  astrometric experiments are included in the IERS list.
\item The characteristics of some sources in the IERS list
  are outdated or doubtful.
\end{itemize}

To overcome this problems, we performed a compilation of new list
of the physical characteristics of geodetic radio sources using the
latest information.
In this paper we present our result of the first stage of this work.

The list of radio sources with their positions was
taken from the Goddard VLBI astrometric
catalogs\footnote{http://vlbi.gsfc.nasa.gov/solutions/astro}, version
2007c, with removing duplicate source 1616+85A (L.Petrov, private
communication) and addition of two ICRF-Ext.2 \cite{Fey04} sources
1039-474 and 1329-665 not included in the Goddard catalog.
It provides 3914 radio sources in total.

At this stage mainly the NASA/IPAC Extragalactic Database
(NED)\footnote{http://nedwww.ipac.caltech.edu/} was scoured.
Some sources were checked with the CfA-Arizona Space
Telescope LEns Survey
(CASTLES)\footnote{http://cfa-www.harvard.edu/glensdata/}
and the HyperLeda\footnote{http://leda.univ-lyon1.fr/} databases.
In this list we have included only the
optical characteristics of geodetic radio sources: source type,
redshift and visual magnitude.
The flux parameters are not included in our list because
they are available from other centers.

\section{List Description}

Our primary interest is to get redshift (z) for astrometric radio
sources to develop the previous researches \cite{MacMillan05,Titov2007}.
In those papers, z values were taken from the ICRF list
\cite{Archival97}.  However, as rather tiny effects in the source
motions are to be investigated, it is important to increase the
number of sources involved in the processing.  Searching the latest
astrophysical databases, primarily the NED, we could considerably
augment the list of geodetic radio sources with known z.
Nevertheless, more than half of the geodetic radio sources have
no determined redshift.

Evidently, the only direct way to get the redshift for other
most frequently observed geodetic sources is to organize a dedicated
observing program with large optical telescopes.  To help in
preparation of such a program, we also collect the source type and
its visual magnitude if this information is available.  Also, it
makes a sense to include in this observational program those sources
with existing but uncertain redshift values.

It should be noted, that not all geodetic radio sources were reliably
identified in the NED. We use the following procedure for
identification.
In the first step, we search for sources by source name
using 'ICRF' and 'IVS' prefix.  So, we rely on the source
identification used in the literature and by the NED staff.  Then
about 500 sources, mostly from the VCS6 list, were searched by
position.  We take into account both the angular distance between the
VLBI and NED positions as well as their uncertainty in
the VLBI and NED positions. For some sources multiply NED objects within the
error level were found.  For 16 sources no appropriate
object was found in the NED, which is mentioned in the comments.
The problem of the source identification in the NED and
other astrophysical data bases hopefully will be solved after
official publication of the VCS6 catalog.

\section{Statistics}

The overall statistics of the new list is the following.

\vskip 2ex
\noindent
\tabcolsep=3pt
\def\arraystretch{0.9}
\begin{tabular}{lrr}
\underline{Number of sources:} && \\
total &&                      3914 (100\%~) \\
\quad ICRF   &  717 (18.3\%) \\
\quad N      & 2376 (60.7\%) \\
\quad S      & 1538 (39.3\%) \\
with known type &&            2369 (60.5\%) \\
\quad AGN    & 1581 (66.7\%) & \\
\quad galaxy &  461 (19.5\%) & \\
\quad other  &  327 (13.8\%) & \\
with known redshift &&        1790  (45.7\%) \\
\quad $<=1$  &  825 (46.1\%) & \\
\quad $>1$   &  965 (53.9\%) & \\
\quad N      & 1185 (66.2\%) & \\
\quad S      &  605 (33.8\%) & \\
with known visual magnitude && 2300 (58.8\%) \\
with known both z and magnitude  && 1739 (44.4\%) \\
with known z or magnitude   && 2351 (60.1\%) \\
with known magnitude and unknown z     &&  561 (14.3\%) \\
without characteristics     && 1563 (39.9\%) \\
\end{tabular}

\begin{figure}[ht!]
\centering
\epsfclipon \epsfxsize=75mm \epsffile{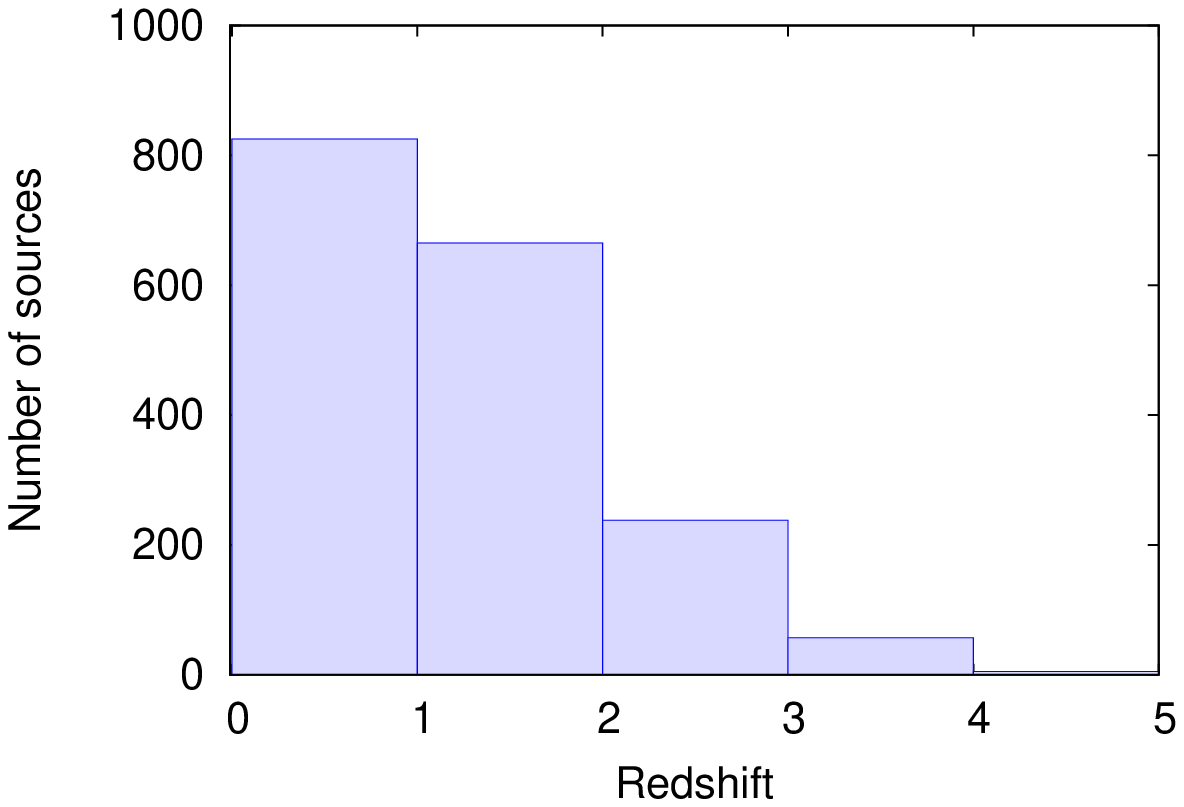}
\hspace{5mm}
\epsfclipon \epsfxsize=75mm \epsffile{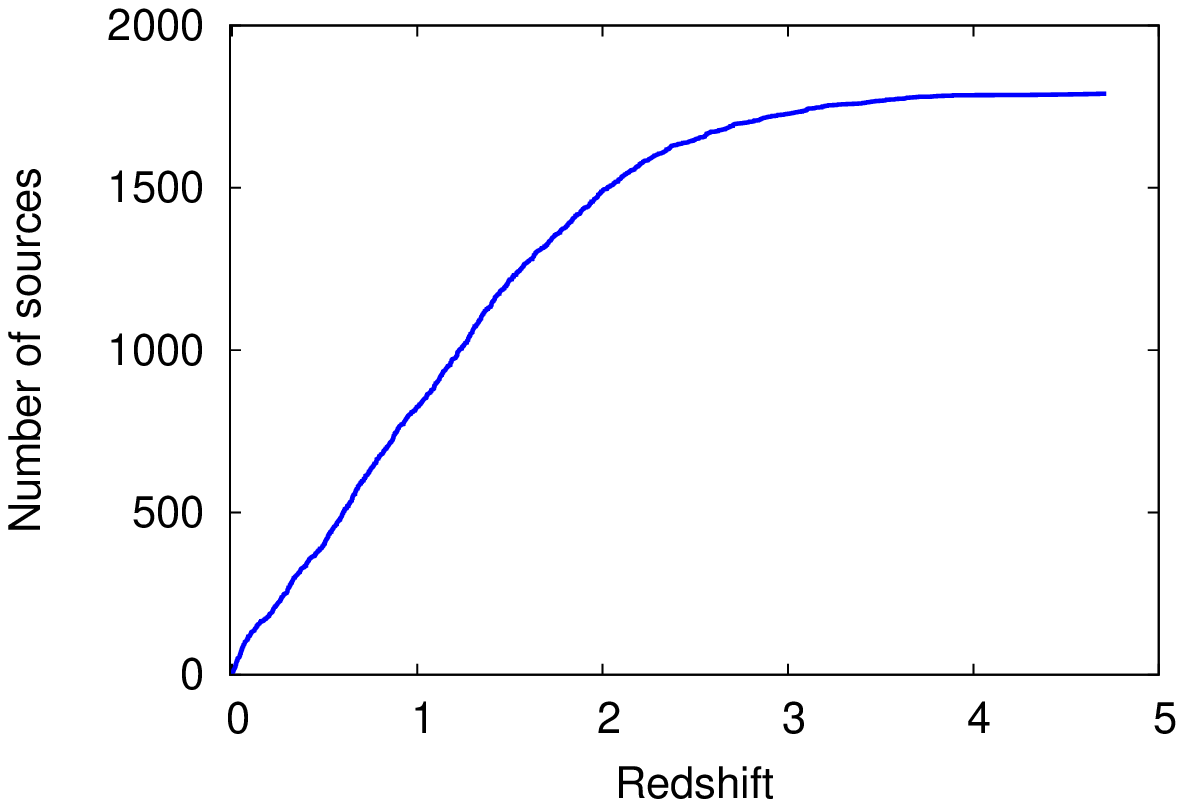}
\caption{Distribution of the redshift ({\it left}) and cumulative
  number of sources ({\it right}).}
\label{fig:z_de_hist}
\end{figure}

\begin{figure}[ht!]
\centering
\epsfclipon \epsfxsize=120mm \epsffile{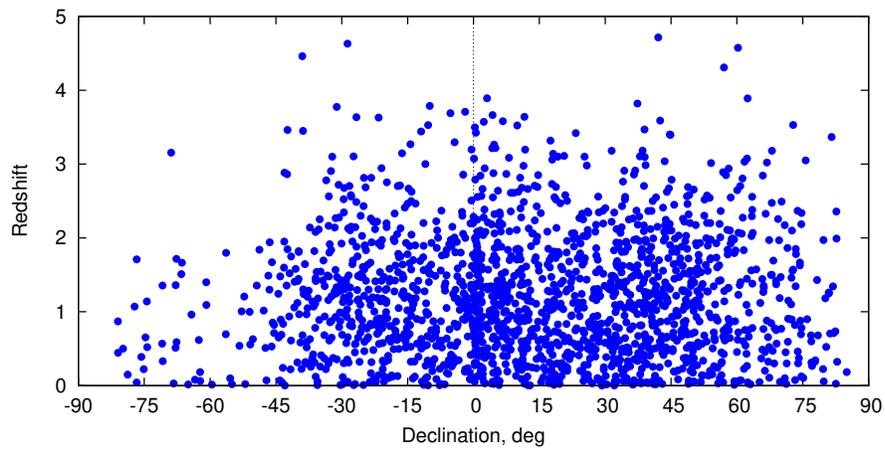}
\caption{Distribution of the redshift over declination.}
\label{fig:z_de_cum}
\end{figure}

Figures \ref{fig:z_de_hist} and \ref{fig:z_de_cum} show the distribution
of the sources with known redshift.

\begin{figure}[ht!]
\centering
\epsfclipon \epsfxsize=75mm \epsffile{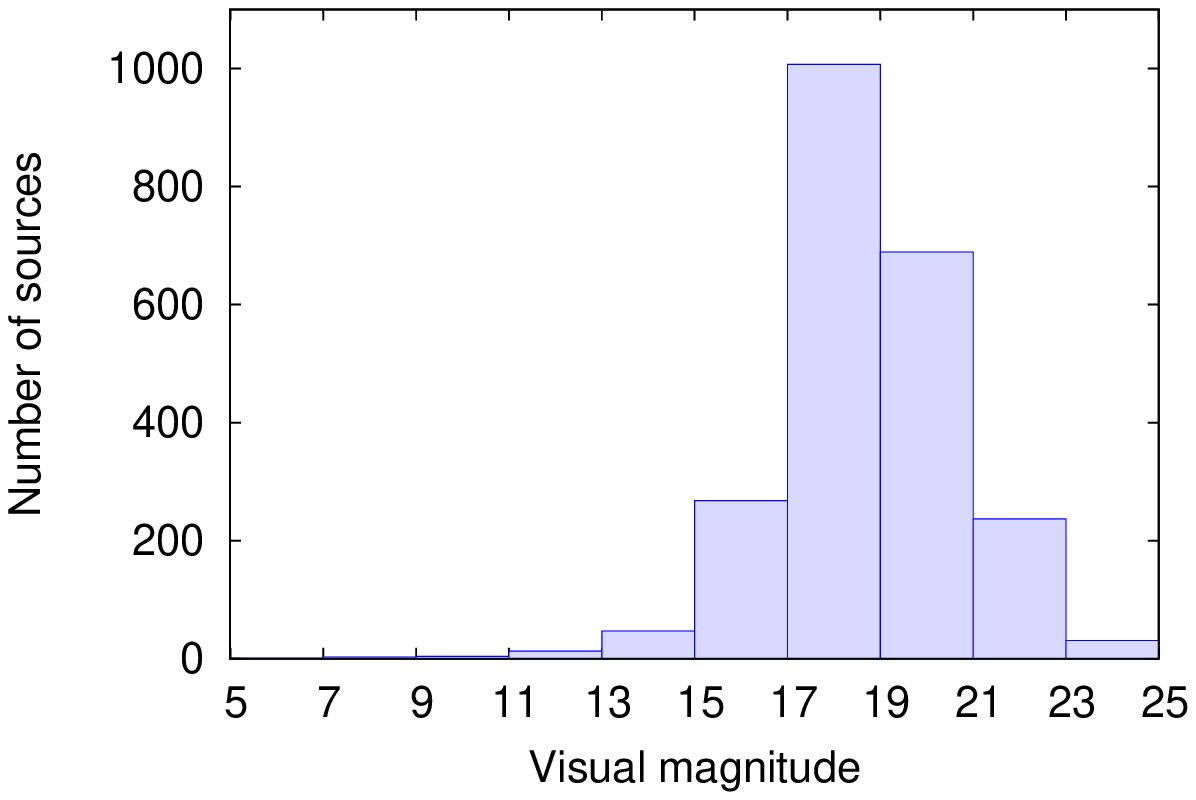}
\hspace{5mm}
\epsfclipon \epsfxsize=75mm \epsffile{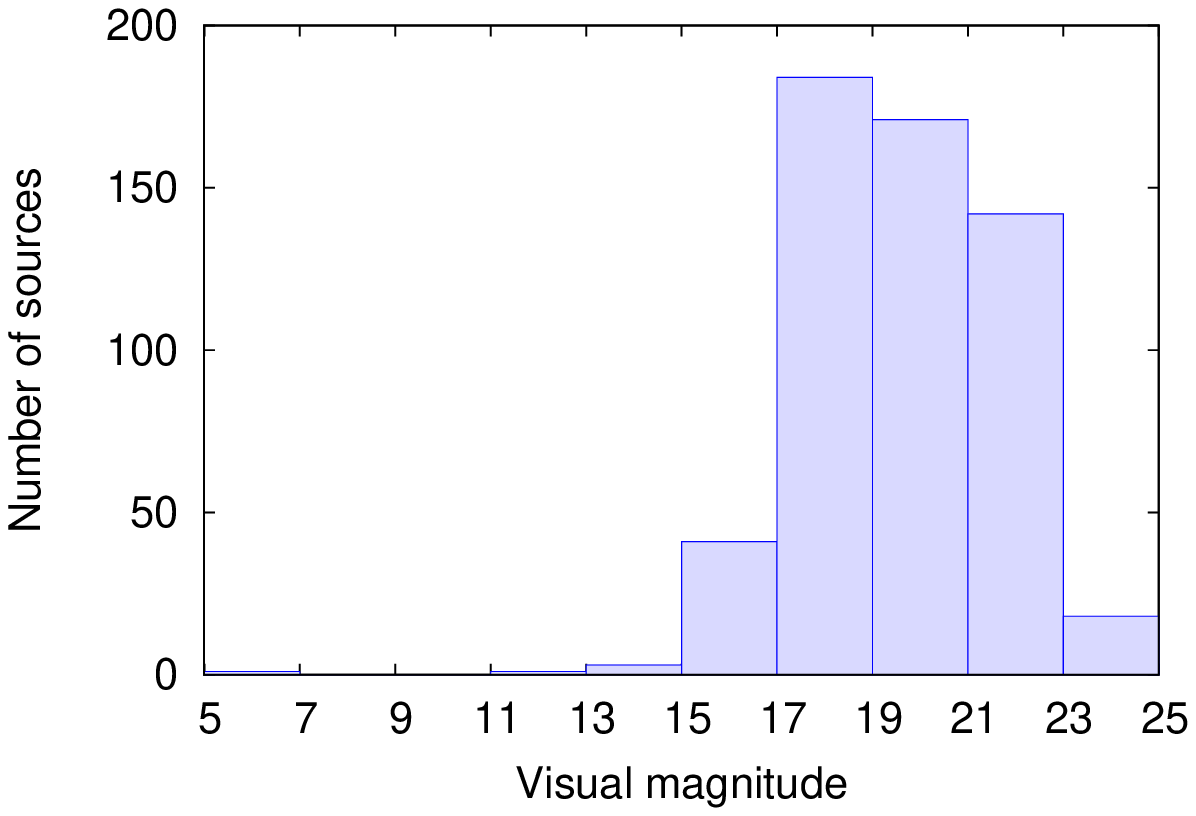}
\caption{Distribution of the visual magnitude for all sources ({\it left})
and for sources without known redshift ({\it right}).}
\label{fig:v_hist}
\end{figure}

Figure \ref{fig:v_hist} shows the distribution of the visual magnitude.
The right part of the figure gives an impression about the magnitude
of the sources for which redshift yet is not determined.

\clearpage    

\section{Comparison with the IERS List}

We have compared our list with the IERS list. 667 common sources were
found. All the sources are present in the IERS list.
Comparison of these two lists results in rather large discrepancy.
\begin{itemize}
\itemsep=-0.3ex
\item The first evident difference is in the number of radio sources.
Our list contains 40 extra ICRF sources plus several hundreds other
sources, 3914 vs. 667 objects in total, 2351 vs. 555 objects with
known z or visual magnitude.
\item
Unlike the IERS list, we did not try to trace all the details of the
Active Galactic Nuclei (AGN) classification that are not always
stable and unambiguous.  So, all the quasars and BL object are
designated as AGN.
\item Redshifts for 55 more ICRF sources were found; redshifts
for 4 sources presented in the IERS list were not included in our list
for various reasons; for 30 sources redshift differs more than
by 0.01; the largest differences are 1.26 (1903-802), 1.20
(1600+431), 0.70 (0646-306).
\item Visual magnitudes for 70 more ICRF sources were found;
magnitudes for 2 sources were not confirmed in our list; for 195
sources magnitude differs more than by 0.5; the largest
differences are 5.2 (1758-651), 5.0 (1156-094, 1322-427), 3.9 (0241+622).
\end{itemize}

Further investigation has to be made to clarify all found
discrepancies between two lists.  It should be mentioned that we
consider as important and useful for a user to provide a detailed
comment in case of doubtful or ambiguous published data.

\section{Conclusion}

A new list of the optical characteristics of geodetic radio sources
has been compiled and available at

\centerline{\tt http://www.gao.spb.ru/english/as/ac\_vlbi/sou\_car.dat.}

This is only the first stage of our work. We are planning the
following steps:
\begin{itemize}
\itemsep=-0.3ex
\item To continue searching for the missing and checking out the
ambiguous characteristics through literature and astronomical
databases.
\item To organize photometric and spectroscopy observations of geodetic
radio sources with large optical telescopes.  In particular, such
observational program has been included in the plan of the Pulkovo
Observatory for 2008. The application for
observation time on the Russian 6-meter BTA telescope for the second
half of 2008 was handed over in February in cooperation with Pulkovo
astrophysicists Kirill Maslennikov and Alexandra Boldycheva.
\end{itemize}

The authors would be happy to know whether this new list is useful
either as a database for VLBI data analysts or as a supplement
material for the ICRF-2 compiling.  We hope that this work will
continue in cooperation with other interested groups.

\section*{Acknowledgements}
This research has made use of the NASA/IPAC Extragalactic Data\-base (NED)
which is operated by the Jet Propulsion Laboratory, California Institute
of Technology, under contract with the National Aeronautics and Space
Administration, the SIMBAD database, operated at CDS, Strasbourg, France,
the HyperLeda database and the CASTLES survey.

\end{document}